\documentclass[11pt]{article}
\usepackage{graphicx}
\usepackage{longtable,lscape}
\usepackage{amsthm}
\usepackage{amsfonts}
\usepackage{amsmath}
\usepackage{amscd}
\usepackage{mathrsfs}
\usepackage{latexsym}
\usepackage{amssymb}
\usepackage{bbm}
\usepackage{float}
\usepackage{url}

\newcommand{\captionfonts}{\footnotesize}
\makeatletter  
\long\def\@makecaption#1#2{%
  \vskip\abovecaptionskip
  \sbox\@tempboxa{{\captionfonts #1: #2}}%
  \ifdim \wd\@tempboxa >\hsize
    {\captionfonts #1: #2\par}
  \else
    \hbox to\hsize{\hfil\box\@tempboxa\hfil}%
  \fi
  \vskip\belowcaptionskip}
\makeatother 

\title{From Ambiguity Aversion to a Generalized Expected Utility. Modeling Preferences in a Quantum Probabilistic Framework}
\author{Diederik Aerts\footnote{Center Leo Apostel for Interdisciplinary Studies and Department of Mathematics, Brussels Free University, Krijgskundestraat 33, 1160 Brussels (Belgium). Email address: \emph{diraerts@vub.ac.be}}  \, and \, Sandro Sozzo\footnote{School of Management and Institute IQSCS, University Road, LE1 7RH Leicester (United Kingdom). Email address: \emph{ss831@le.ac.uk}}}

\date{}

\begin{document}

\maketitle

\begin{abstract}
\noindent
Ambiguity and ambiguity aversion have been widely studied in decision theory and economics both at a theoretical and an experimental level. After Ellsberg's seminal studies challenging subjective expected utility theory (SEUT), several (mainly normative) approaches have been put forward to reproduce ambiguity aversion and Ellsberg-type preferences. However, Machina and other authors have pointed out some fundamental difficulties of these generalizations of SEUT to cope with some variants of Ellsberg's thought experiments, which has recently been experimentally confirmed. Starting from our quantum modeling approach to human cognition, we develop here a general probabilistic framework to model human decisions under uncertainty. We show that our quantum theoretical model faithfully represents different sets of data collected on both the Ellsberg and the Machina paradox situations, and is flexible enough to describe different subjective attitudes with respect to ambiguity. Our approach opens the way toward a quantum-based generalization of expected utility theory (QEUT), where subjective probabilities depend on the state of the conceptual entity at play and its interaction with the decision-maker, while preferences between acts are determined by the maximization of this `state-dependent expected utility'.
\end{abstract}

\section{Introduction\label{intro}}
The deviations from traditional probabilistic models of human cognition that have been experimentally detected up to now can be roughly divided into two groups, `probability judgment errors' (e.g., `conjunction and disjunction fallacy' (Tversky \& Kahneman, 1983; Morier \& Borgida, 1984), `concept categorization' (Hampton, 1988a,b), `unpacking effects' (Fox \& Tversky, 1998) and `decision-making errors' (e.g., `disjunction effect' (Tversky \& Shafir, 1992), `Allais, Ellsberg and Machina paradoxes' (Allais, 1953; Ellsbeg, 1961; Machina, 2009)). These traditional models of cognition rest on a classical vision of probability theory, which can be traced back to the axiomatization proposed by Kolmogorov in the thirties (Kolmogorov, 1933). When applied to decision theory and economics, this traditional approach to probability has produced a unified normative axiomatic theory to model human decisions under uncertainty, which has been predictively successful for years since the forties, ` expected utility theory' (EUT) (von Neumann \& Morgenstern, 1944; Savage, 1954). By following Knight's original distinction of the different forms of uncertainty (Knight, 1921), one nowadays distinguishes between `risk', that is, `uncertainty about known probabilities', and `ambiguity', that is, `uncertainty about unknown probabilities'. And also EUT takes a different mathematical formulation, depending on whether situations in which risk is present, `objective expected utility theory' (OEUT) (von Neumann \& Morgenstern, 1944), or   situations in which ambiguity is present, `subjective expected utility theory' (SEUT) (Savage, 1954). It should however be noticed that the latter situations have manifold implications in real-life decisions, from financial asset pricing to marketing choices, portfolio management, medical treatment decisions, as subjective probabilities frequently appear in everyday decisions, in the form of `beliefs about likelihood of events'.

The first experimental challenge to OEUT was identified by Alain Allais in the fifties (Allais, 1953). Coming to SEUT, its drawbacks arose instead with the so-called `Ellsberg urns' (Ellsberg, 1961). Daniel Ellsberg presented in 1961 a series of thought experiments with urns and balls of different colors, e.g., `two-color example' and `three-color example', where he predicted that real human decisions would have refuted the predictions of SEUT. Ellsberg suggested that people prefer to take actions associated with events over known rather unknown probabilities or, in other words, `people prefer known versus unknown probabilities'. This conservative attitude was called `ambiguity aversion' by Ellsberg, and it violates a specific axiom of SEUT, the `Sure-Thing principle' -- a violation of this principle is also observed in other decision-making errors, such as the disjunction effect (Tversky \& Shafir, 1992). Ambiguity aversion has been systematically confirmed by several cognitive experiments collected in the last forty years (see, e.g., (McCrimmon \& Larsson, 1979; Einhorn \& Hogarth, 1986; Camerer \& Weber, 1992; Fox \& Tversky, 1995); see also the exhaustive review by Machina and Siniscalchi (2014)), but the one by Slovic and Tversky (1974), who found an `ambiguity seeking' attitude. Both ambiguity aversion and ambiguity attraction are however incompatible with the predictions of SEUT, which led many scholars to look for theoretical alternatives that could accomodate the behavior observed in Ellsberg experiments. These alternatives have generally a normative status, that is, they describe what people should not, not what people actually do, and have different names and scopes, for example, `expected utility with multiple priors' (Gilboa \& Schmeidler, 1989), `Choquet expected utility theory' (CEUT) (Schmeidler, 1989), `smooth ambiguity preferences model' (Klibanoff, Marinacci \& Mukerij, 2005), `variational preference model' (Maccheroni, Marinacci \& Rustichini, 2006), and `cumulative prospect theory (Tversky \& Kahneman, 1992), within Tversky-Kahneman theory of human heuristics and bias (Tversky \& Kahneman, 1974).

These generalizations of SEUT were seriously challenged by two thought experiments recently presented by Mark Machina (2009), the `50:51 example' and the `reflection example' (Machina 2009; Baillon, L'Haridon \& Placido, 2011). In particular, Machina introduced a new element, `informational symmetry', which violates an axiom of CEUT, called `tail separability', exactly as ambiguity aversion violates the Sure-Thing principle. Without entering into the technical details of the violation (some aspects of it will be presented in Section \ref{ellsbergmachina}), we want to emphasize an implicit assumption of SEUT that is weakened in the above generalizations. SEUT assumes that a `single Kolmogorovian probability distribution over a single $\sigma$-algebra of events is defined which models subjective probabilities' in human decisions. Departure from this assumption will become crucial in our approach, as we will see in Section \ref{quantummodels}.

We have worked in the last years on the identification of quantum structures in cognition and the mathematical modeling of decision-making under uncertainty, obtaining significant results in the explanation of the so-called human probability judgment errors in terms of genuine quantum aspects (emergence, entanglement, indistinguishability, interference, superposition)
(Aerts \& Gabora, 2005a,b; Aerts, 2009a; Aerts, Broekaert, Gabora \& Sozzo, 2013; Aerts, Gabora \& Sozzo, 2013; Aerts, Sozzo \& Veloz, 2015a,b; see also Pothos \& Busemeyer, 2009; Busemeyer \& Bruza, 2012; Pothos \& Busemeyer, 2013; Wang, Solloway, Shiffrin \& Busemeyer, 2014). For what explicitly concerns ambiguity and ambiguity aversion, we have worked out a quantum theoretical framework to model both the Ellsberg and the Machina paradox situations (Aerts, Sozzo \& Tapia, 2012, 2014). Our approach rests on two fundamental observations, as follows. 

(i) In an Ellsberg/Machina-type decision-making process, the agent's choice is actualized as a consequence of an interaction with the cognitive context, exactly as in a quantum measurement process the measurement outcome is actualized as a consequence of the interaction of the measured particle with the measuring apparatus. Therefore, in cognitive entities, as well as in microscopic quantum entities, measurements do not reveal preexisting values of the observed properties but, rather, they actualize genuine potentialities. Kolmogorovian probability can only formalize lack of knowledge about actualities, hence it is generally not able to cope with such a decision-making process.  We have proven that this is possible in Ellsberg/Machina-type decisions by using a complex Hilbert space and representing probability measures by means of `projection valued measures' on this complex Hilbert space. A projection valued measure is essentially different from a single Kolmogorovian probability measure, since the latter is a $\sigma $-algebra valued measure, whilst the former is not, due to lack of distributivity (Aerts, Sozzo \& Tapia, 2012). 

(ii) The above notion of ambiguity is completely compatible, both at a mathematical and an intuitive level, with the representation of states of cognitive entities as vectors of a Hilbert space. Indeed, just like in quantum theory the state vector incorporates the `quantum uncertainty' of a microscopic particle, also in an Ellsberg/Machina-type situation, the agent's subjective preference toward ambiguity is naturally formalized by representing the conceptual situation the agent is confronted with by means
of such a Hilbert space vector. In our approach, ambiguity aversion is only one of the conceptual landscapes surrounding the decision-maker's choice in a situation where ambiguity is present (Aerts, Sozzo \& Tapia, 2014). This representation is compatible with the experimental findings confirming Ellsberg's prediction about the human attitude toward ambiguity, but also with some recent experiments where such attitude is more controversial (see, e.g., (Slovic \& Tversky, 1974; Binmore, Stewart \& Voorhoeve, 2012); see also the review (Machina \& Siniscalchi, 2014)).

We develop in Section \ref{quantummodels} an amended and updated theoretical framework, where subjective probabilities and preferences in the Ellsberg and Machina paradox situations are modeled by using the mathematical formalism of quantum theory, which is briefly summarized in \ref{quantum}. Further, we show in the same section that our quantum theoretical model successfully represents both an experiment on the Ellsberg three-color example by ourselves, confirming Ellsberg-type preferences (Aerts, Sozzo \& Tapia, 2012), and an experiment on the Machina reflection example by l'Haridon and Placido (2010), confirming Machina-type preferences.

Our quantum theoretical framework opens the way to a `contextual quantum-based generalization of EUT' (QEUT), where human preferences also depend on the conceptual, not only the physical, state encoding the potential effects of the cognitive context. These effects include, in particular, subjective preferences toward ambiguity and ambiguity aversion. Because of Gleason theorem (Gleason, 1957), this conceptual state is bijectively associated with a standard quantum probability distribution modeling the subjective probabilities of the decision-maker in that conceptual state. Whenever the conceptual state changes as a consequence of, e.g., a decision, also the quantum probability distribution modeling the subjective probability changes, which enables representation of Ellsberg-type and Machina-type situations. Therefore, each act is associated with a `state-dependent expected utility' in QEUT, and human preferences are determined by the maximization of this state-dependent expected utility. This change of perspective is important, in our opinion, because it enables modeling of different types of human behavior, from ambiguity aversion to ambiguity attraction, Sure-Thing principle, principle of insufficient reason, as we conclude in Section \ref{QEUT}.

\section{Expected utility, Ellsberg and Machina paradoxes\label{ellsbergmachina}}
In this section we summarize the essential definitions and results of SEUT that we need to discuss Ellsberg and Machina paradoxes (Savage, 1954). Our basic mathematical framework requires a set ${\mathscr S}$ of physical states, a $\sigma$-algebra ${\mathscr A}\subseteq {\mathscr P}({\mathscr S})$ of subsets of ${\mathscr S}$ which we call `events', and a subjective probability measure $p:{\mathscr A}\subseteq {\mathscr P}({\mathscr S})\longrightarrow [0,1]$ over $\mathscr A$. Then, we denote by ${\mathscr X}$ the set of all consequences. An `act' is defined as a function $f: {\mathscr S} \longrightarrow {\mathscr X}$. Next, we introduce a weak preference relation $\succsim$ over acts, where $\succ$ and $\sim$ denote strong preference and indifference, respectively. Finally, let $u: {\mathscr X} \longrightarrow \Re$ denote a `Bernoulli utility function' over the set of consequences, and suppose that $u$ is strictly increasing and continuous. 

For the sake of simplicity, let us introduce the following simplifications. Firstly, let us only consider discrete and finite cases. Secondly, let us introduce the set $\{ E_{1}, E_{2}, \ldots, E_{n}  \}$ of events, and suppose that $E_{1}, E_{2}, \ldots, E_{n}$ are mutually exclusive and exhaustive, i.e. form a partition of $\mathscr S$. Thirdly, assume that, for every $i=1,2,\ldots,n$, the event $E_{i}$ occurs with probability $p_i$, which might be unknown. Let $f$ be the act that, for every  $i=1,2,\ldots,n$, associates the event $E_{i}$ with the consequence $x_{i}$, and suppose that $x_i\in \Re$, so that $x_i$ denotes a monetary payoff. Then, the act $f$ can be represented as the 2n-ple $f=(E_1,x_1;E_2,x_2;\ldots;E_n,x_n)$. As usual, we define the expected utility of the act $f$ by means of the functional $W(f)=W(E_1,x_1;E_2,x_2;\ldots;E_n,x_n)=\sum_{i=1}^{n} p_i u(x_i)$.

Let now $f$ and $g$ be two acts, and let $W(f)$ and $W(g)$ be the corresponding expected utilities. Then, Savage proved that, if some axioms are satisfied, including the Sure-Thing principle, we have $f \succsim g$ if and only if $W(f)\ge W(g)$ (Savage, 1954). We stress that a single subjective probability distribution $p$ is assumed which is defined over a single $\sigma$-algebra ${\mathscr A}$ of events.

However, in 1961 Ellsberg firstly predicted in his `two-color example' that people do not always choose by maximizing their subjective expected utility, but they generally prefer acts over events with known (or objective) probabilities to acts over events with unknown (or subjective) probabilities, a phenomenon called `ambiguity aversion'. The thought experiment where ambiguity aversion manifestly clashes with SEUT is the `three-color example'. 

Consider one urn with 30 red balls and 60 balls that are either yellow or black in unknown proportion. One ball will be drawn at random from the urn. Then, free of charge, a person is asked to bet on one of the acts $f_1$, $f_2$, $f_3$ and $f_4$ defined in Table 1. 
\noindent 
\begin{table} \label{table01}
\begin{center}
\begin{tabular}{|p{1.5cm}|p{1.5cm}|p{1.5cm}|p{1.5cm}|}
\hline
Act & Red & Yellow & Black \\ 
\hline
\hline
$f_1$ & \$100 & \$0 & \$0 \\ 
\hline
$f_2$ & \$0 & \$0 & \$100 \\ 
\hline
$f_3$ & \$100 & \$100 & \$0 \\ 
\hline
$f_4$ & \$0 & \$100 & \$100 \\ 
\hline
\end{tabular}
\end{center}
{\bf Table 1.} The payoff matrix for the Ellsberg three-color thought experiment.
\end{table}
\noindent 
Ellsberg suggested that, when asked to rank these acts, most of the persons will prefer $f_1$ over $f_2$ and $f_4$ over $f_3$. On the other hand, acts $f_1$ and $f_4$ are `unambiguous', as they are associated with events over known probabilities, while 
acts $f_2$ and $f_3$ are `ambiguous', as they are associated with events over unknown probabilities. Conclusion follows at once. People prefer the unambiguous act over its ambiguous counterpart. There is a huge experimental evidence that decision-makers actually show this ambiguity aversion (see, e.g., the extensive review by Machina and Siniscalchi (2014)). Only the experiment by Slovic and Tversky (1974) indicated that `ambiguity attraction' was at play, while more recent experiments identify other behavioral mechanisms as primary (Binmore, Stewart \& Voorhoeve, 2012).

Neither ambiguity aversion nor ambiguity attraction can be explained within SEUT, as they violate the Sure-Thing principle, according to which, preferences should be independent of the common outcome. In the specific case of the three-color urn, preferences should not depend on whether the common event ``a yellow ball is drawn'' pays off \$0 or \$100. More technically, SEUT predicts `consistency of decision-makers' preferences', that is, $f_1$ is preferred to $f_2$ if and only if $f_3$ is preferred to $f_4$. A simple calculation shows that this is impossible. Indeed, if we denote by $p_{R}$, $p_{Y}$ and $p_{B}$ the subjective probability that a red ball, a yellow ball, a black ball, respectively, is drawn (with $p_{R}=1/3=1-(p_{Y}+p_{B})$), then the expected utilities $W(f_{i})$, $i=1,2,3,4$, are such that
$W(f_1)>W(f_2)$ if and only if $(p_{R}-p_{B})(u(100)-u(0))>0$ if and only if $W(f_3)>W(f_4)$. Hence, no assignment of the subjective probabilities $p_{R}$, $p_{Y}$ and $p_{B}$ reproduces a preference with $W(f_1)>W(f_2)$  and $W(f_4)>W(f_3)$. 

In the last forty years several generalizations of SEUT have been proposed in order to account for ambiguity and/or ambiguity aversion, introducing expected utility with multiple priors (Gilboa \& Schmeidler, 1989), Choquet expected utility (Schmeidler, 1989), smooth ambiguity preferences (Klibanoff, Marinacci \& Mukerij, 2005), variational preferences (Maccheroni, Marinacci \& Rustichini, 2006), etc. (Machina \& Siniscalchi, 2014). In particular, CEUT, which ranks events each other,
has been quite successful to model Allais and Ellsberg paradoxes, and in many applications in economics and finance (see, e.g., (Hansen \& Sargent, 2000)). However, in 2009 Mark Machina proposed new thought experiments, which seriously challenged these normative approaches. We have discussed in detail the `50:51 example' in (Aerts, Sozzo \& Tapia, 2012). Here we analyze the `reflection example' (Machina, 2009; Baillon, l'Haridon \& Placido, 2011), which questions an axiom of CEUT, the so-called `tail separability', exactly as the Ellsberg three-color example questions the Sure-Thing principle of SEUT.

It is possible to elaborate two versions of the reflection example, as follows (l'Haridon \& Placido, 2010).

`Reflection example with lower tail shifts'.  Consider one urn with 20 balls, 10 are either red or yellow in unknown proportion, 10 are either black or green in unknown proportion. One ball will be drawn at random from the urn. Then, free of charge, a person is asked to bet on one of the acts $f_1$, $f_2$, $f_3$ and $f_4$ defined in Table 2. 
\noindent 
\begin{table} \label{table02}
\begin{center}
\begin{tabular}{|p{1.5cm}|p{1.5cm}|p{1.5cm}|p{1.5cm}|p{1.5cm}|}
\hline
Act & Red & Yellow & Black & Green \\ 
\hline
\hline
$f_1$ & \$0 & \$50 & \$25 & \$25 \\ 
\hline
$f_2$ & \$0 & \$25 & \$50 & \$25 \\ 
\hline
$f_3$ & \$25 & \$50 & \$25 & \$0 \\ 
\hline
$f_4$ & \$25 & \$25 & \$50 & \$0 \\ 
\hline
\end{tabular}
\end{center}
{\bf Table 2.} The payoff matrix for the Machina reflection example with lower tail shifts.
\end{table}
\noindent 
Machina introduced the notion of `informational symmetry', namely, the events ``the drawn ball is red or yellow'' and ``the drawn ball is black or green'' have known and equal probability and, further, the ambiguity about the distribution of
colors is similar in the two events. In an informational symmetry perspective, people should prefer act $f_1$ over act $f_2$ and act $f_4$ over act $f_3$, or they should prefer act $f_2$ over act $f_1$ and act $f_3$ over act $f_4$. On the other hand, let us introduce the utilities $u(0)$, $u(25)$ and $u(50)$, the subjective probabilities $p_R$, $p_Y$, $p_B$ and $p_G$, and calculate the expected utilities $W(f_i)$ of the acts $f_i$, $i=1,2,3,4$. Then, according to SEUT, we have again that preferences should be consistent, namely, $W(f_1)>W(f_2)$
 if and only if $(u(50)-u(25))(p_Y-p_B)>0$ if and only if $W(f_3)>W(f_4)$. The interesting aspect of this example is that CEUT predicts 
 similar consistency requirements on the basis of tail separability. A recent experiment by L'Haridon and Placido (2010) confirms instead the Machina preference $f_1 \succsim f_2$ and  $f_4 \succsim f_3$, consistently with informational symmetry. We will discuss the details of this experiment in Section \ref{quantummodels}, where we will represent it within our general quantum theoretical modeling. But, we can already maintain that such type of preferences needs a completely new theoretical approach to ambiguity.

`Reflection example with upper tail shifts'.  Consider one urn with 20 balls, 10 are either red or yellow in unknown proportion, 10 are either black or green in unknown proportion. One ball will be drawn at random from the urn. Then, free of charge, a person is asked to bet on one of the acts $f_1$, $f_2$, $f_3$ and $f_4$ defined in Table 3. 
\noindent 
\begin{table} \label{table03}
\begin{center}
\begin{tabular}{|p{1.5cm}|p{1.5cm}|p{1.5cm}|p{1.5cm}|p{1.5cm}|}
\hline
Act & Red & Yellow & Black & Green \\ 
\hline
\hline
$f_1$ & \$50 & \$50 & \$25 & \$75 \\ 
\hline
$f_2$ & \$50 & \$25 & \$50 & \$75 \\ 
\hline
$f_3$ & \$75 & \$50 & \$25 & \$50 \\ 
\hline
$f_4$ & \$75 & \$25 & \$50 & \$50 \\ 
\hline
\end{tabular}
\end{center}
{\bf Table 3.} The payoff matrix for the Machina reflection example with upper tail shifts.
\end{table}
\noindent 
According to Machina's informational symmetry, people should again prefer act $f_1$ over act $f_2$ and act $f_4$ over act $f_3$, or they should prefer act $f_2$ over act $f_1$ and act $f_3$ over act $f_4$. On the other hand, let us introduce the utilities $u(25)$, $u(50)$ and $u(75)$, the subjective probabilities $p_R$, $p_Y$, $p_B$ and $p_G$, and calculate the expected utilities $W(f_i)$ of the acts $f_i$, $i=1,2,3,4$. Then, according to SEUT, we have that preferences should again be consistent, namely, $W(f_1)>W(f_2)$
 if and only if $(u(50)-u(25))(p_Y-p_B)>0$ if and only if $W(f_3)>W(f_4)$. One shows that tail separability of CEUT leads to 
 a similar prediction. The experiment by L'Haridon and Placido (2010) confirms again the Machina preference $f_1 \succsim f_2$ and  $f_4 \succsim f_3$, consistently with informational symmetry. Necessity of new approaches to model this type of preferences follows at once.

\section{A quantum theoretical framework to model human preferences\label{quantummodels}}
We develop in this section an amended and more rigorous 
theoretical framework to model Ellsberg and Machina  paradox situations that uses the mathematical formalism of quantum theory, whose 
essentials 
are summarized in  \ref{quantum}. Our approach rests on the results in (Aerts, Sozzo \& Tapia, 2012, 2014). The fundamental assumptions that we make are the following.

(i) The conceptual (Ellsberg, Machina, etc.) situation is
described by a well defined 
state $p_{v}$. This state has a conceptual nature, hence it should be distinguished from a physical state (see Section \ref{ellsbergmachina}). The conceptual state mathematically captures aspects of ambiguity.

(ii) There is a contextual interaction of a cognitive, not physical, nature between the decision-maker and the conceptual situation that is the object of the decision. This contextual interaction determines a change of the state of the conceptual entity.

(iii) The conceptual situation, state and decision-making process are modeled by using the mathematical formalism of quantum theory. It follows from Gleason theorem (Gleason, 1957) that the conceptual state identifies a single quantum probability distribution through the Born rule and, for some states, we can interpret this probability distribution as subjective probability for the decision-making process.

Let us now consider the Ellsberg three-color example and represent it by using the quantum mechanical formalism. The conceptual Ellsberg entity is the urn with 30 red balls and 60 yellow and black balls in unknown proportion. 
Since we have introduced the notion of state to describe the conceptual entity, we denote by $p_v$ 
a state describing 
a possible `conceptual state of being' of the Ellsberg entity with 30 red balls and 60 yellow and black balls in unknown proportion. 
Note that several `conceptual states of being' of the 
Ellsberg entity are possible, if we, for example, 
also specify something more than just `unknown proportion' about yellow and black balls, and then this different conceptual situation of the Ellsberg entity would be represented by a different state.
Because we use the quantum formalism 
each state $p_v$ is represented by the unit vector $|v\rangle$ of the complex Hilbert space ${\mathbb C}^{3}$ over complex numbers. The choice of ${\mathbb C}^{3}$ is due to the fact that we have three mutually exclusive and exhaustive events in the three-color example -- the generalization to the Ellsberg n-color example is straightforward. We denote by $(1,0,0)$, $(0,1,0)$ and $(0,0,1)$ the unit vectors of the canonical basis of ${\mathbb C}^{3}$.

Let us consider a `color measurement' $e$ performed on the conceptual Ellsberg entity in a 
state $p_v$. This color measurement has three outcomes $o_{R}$, $o_{Y}$, $o_{B}$, corresponding to the three colors red, yellow and black, respectively, and it is represented by the self-adjoint operator ${\mathscr E}$ with spectral family $\{ P_{i}=|i\rangle\langle i | \}_{i=R,Y,B}$, where $|R\rangle$, $|Y\rangle$ and $|B\rangle$ are the eigenvectors associated with the eigenvalues $o_{R}$, $o_{Y}$ and $o_{B}$, respectively, of ${\mathscr E}$. Equivalently, we have ${\mathscr E}=\sum_{i=1}^{3} o_{i}P_{i}$. We do not lose in generality by setting $|R\rangle=(1,0,0)$, $|Y\rangle=(0,1,0)$,  and $|B\rangle=(0,0,1)$. With this choice, 
a state $p_v$ of the conceptual Ellsberg entity is represented by the unit vector
\begin{equation}
|v \rangle=\rho_R e^{i \theta_{R}}|R\rangle+\rho_Y e^{i \theta_{Y}}|Y\rangle+\rho_B e^{i \theta_{B}}|B\rangle=(\rho_R e^{i \theta_{R}},\rho_Y e^{i \theta_{Y}},\rho_B e^{i \theta_{B}})
\end{equation}
By using quantum probabilistic rules, the probability $\mu_{v}(o_{i}, e)$ of drawing a ball of color $i$, $i=R,Y,B$, when the conceptual Ellsberg entity is in 
a state $p_v$ is given by
\begin{equation}
\mu_{v}(o_{i}, e)=\langle v | P_{i} | v \rangle=|\langle i | v \rangle|^{2}=\rho_{i}^{2}
\end{equation}
We have $\rho_{R}^{2}=1/3$, as the urn contains 30 red balls. Therefore, 
a state $p_v$ of the conceptual Ellsberg entity is represented by the unit vector 
\begin{equation}  \label{ellsbergstate}
|v \rangle=(\frac{1}{\sqrt{3}} e^{i \theta_{R}},\rho_Y e^{i \theta_{Y}}, \sqrt{\frac{2}{3}-\rho_{Y}^{2}} e^{i \theta_{B}})
\end{equation}
Let us now introduce two 
specific states of the conceptual Ellsberg entity. 
The state $p_{RY}$ describes the conceptual situation where there are no black balls, hence it is represented by the unit vector 
\begin{equation} \label{noblack}
|v_{RY} \rangle=(\frac{1}{\sqrt{3}} e^{i \theta_{R}},\sqrt{\frac{2}{3}} e^{i \theta_{Y}}, 0)
\end{equation}
The state $p_{RB}$ describes instead the conceptual situation where there are no yellow balls, hence it is represented by the unit vector 
\begin{equation} \label{noyellow}
|v_{RB} \rangle=(\frac{1}{\sqrt{3}} e^{i \theta_{R}},0, \sqrt{\frac{2}{3}} e^{i \theta_{B}})
\end{equation}

The Ellsberg entity is a conceptual entity, which means that `cognitive contexts' have an influence on its state, and in general will make a specific state change into another state. This is what happens in the conceptual realm in analogy with the physical realm, where a physical context will in general change the physical state of a physical entity. Hence, whenever a decision-maker is asked to ponder between the choice of $f_1$ and $f_2$, the pondering itself, before a choice is made, is a cognitive context, and hence it changes in general the state of the conceptual Ellsberg entity. Equally so, whenever a decision-maker is asked to ponder between the choice of $f_3$ and $f_4$, also this introduces a cognitive context, before the choice is made  -- and a context which in general is different from the one introduced by pondering about the choice between $f_1$ and $f_2$ -- which will in general change the state of the conceptual Ellsberg entity.

Let us now introduce a state $p_0$ describing the situation where no cognitive context is present, and call it the `ground state' of the Ellsberg entity. We can identify this ground state by introducing a new measurement, which would consist of asking a subject to `estimate the number of balls', red, yellow and black, without mentioning any reference to a bet that has to be taken. The red balls will be estimated to be 30 in number, because that is a given amount. For the black and yellow balls subject might estimate differently, but if no conceptual context at all is present, and hence only the presentation of the Ellsberg entity itself is offered to the subjects, on average estimations will be 30 black balls and 30 yellow balls, as a consequence of the symmetry present in the Ellsberg entity -- unknown number of black and yellow their sum being equal to 60 induces the symmetry 30 black and 30 yellow.
More specifically,  the unit vector $|v_0\rangle=\frac{1}{\sqrt{3}}(1,1,1)$ in the canonical basis of ${\mathbb C}^3$ represents well this symmetry, and hence we will use it to describe the ground state $p_0$ of the Ellsberg entity. Then, a pondering about the choice between $f_1$ and $f_2$ will make the state $p_0$ of the conceptual Ellsberg entity change to a state $p_{w_1}$ that is generally different from the state $p_{w_2}$ in which the Ellsberg conceptual entity changes from $p_0$ when pondering about a choice between $f_3$ and $f_4$. In particular, a `highly ambiguity averse' decision-maker will be, as a consequence of his or her pondering in the choice between acts $f_1$ and $f_2$, confronted with the conceptual Ellsberg entity which changes from the 
ground state $p_0$ to a state $p_{w_1}$ that is very close to the state $p_{RY}$ represented in (\ref{noblack}). Analogously, a `highly ambiguity averse' decision-maker will be, 
as a consequence of his or her pondering in the choice between acts $f_3$ and $f_4$, confronted with the conceptual Ellsberg entity which changes from the state $p_0$ to a state $p_w$ that is very close to the state $p_{RB}$ represented in (\ref{noyellow}).

Let us now come to the representation of the acts $f_1$, $f_2$, $f_3$ and $f_4$ in Table 1, Section \ref{ellsbergmachina}. For a given utility function $u$, to be determined experimentally, we respectively associate $f_1$, $f_2$, $f_3$ and $f_4$ with the self-adjoint `utility operators'
\begin{eqnarray}
{\mathscr F}_{1}&=&u(100)P_R+u(0)P_Y+u(0)P_B \label{f1}\\
{\mathscr F}_{2}&=&u(0)P_R+u(0)P_Y+u(100)P_B \label{f2}\\
{\mathscr F}_{3}&=&u(100)P_R+u(100)P_Y+u(0)P_B\label{f3} \\
{\mathscr F}_{4}&=&u(0)P_R+u(100)P_Y+u(100)P_B \label{f4}
\end{eqnarray}
The corresponding expected utilities in 
a state $p_{v}$ of the conceptual Ellsberg entity are 
\begin{eqnarray}
W_{v}(f_1)&=&\langle v| {\mathscr F}_{1}|v\rangle=\frac{1}{3}u(100)+\frac{2}{3}u(0) \label{w1}\\
W_{v}(f_2)&=&\langle v| {\mathscr F}_{2}|v\rangle=(\frac{1}{3}+\rho_Y^2)u(0)+(\frac{2}{3}-\rho_{Y}^{2})u(100) \label{w2} \\
W_{v}(f_3)&=&\langle v| {\mathscr F}_{2}|v\rangle=(\frac{1}{3}+\rho_Y^2)u(100)+(\frac{2}{3}-\rho_{Y}^{2})u(0) \label{w3}\\
W_{v}(f_4)&=&\langle v| {\mathscr F}_{4}|v\rangle=\frac{1}{3}u(0)+\frac{2}{3}u(100) \label{w4}
\end{eqnarray}
As we can see, $W_{v}(f_1)$ and $W_{v}(f_4)$ do not depend on the state $p_v$, and this would be the case for an arbitrary state $p$, hence they are ambiguity-free, i.e. independent of the state, while $W_{v}(f_2)$ and $W_{v}(f_3)$ do depend on $p_v$, and will depend on any state. This means that it is possible to find a state $p_{w_1}$ such that $W_{w_1}(f_1) > W_{w_2}(f_2)$, e.g., the state represented in (\ref{noblack}), and a state $p_{w_2}$ such that $W_{w_2}(f_4) > W_{w_2}(f_3)$, e.g., the state represented in (\ref{noyellow}). These two states perfectly reproduce Ellsberg preferences, in agreement with an ambiguity aversion assumption.

Let us now study how the quantum theoretical model above works in practice. We performed a cognitive experiment in which we asked 57 persons, chosen among our colleagues and friends, to rank the four acts in Table 1, Section \ref{ellsbergmachina}. The details of the experiment are presented in (Aerts, Sozzo \& Tapia, 2014). 
 In it, 34 participants preferred acts $f_1$ and $f_4$, 12 participants preferred acts $f_2$ and $f_3$, 7 participants preferred acts $f_2$ and $f_4$, and 6 participants preferred acts $f_1$ and $f_3$. This makes the weights with preference of acts $f_1$ over act $f_2$ to be 0.68 against 0.32, and the weights with preference of act $f_4$ over act $f_3$ to be 0.69 against 0.31. Hence, 34+12=46 persons over 57 chose $f_1$ and $f_4$ or the inversion $f_2$ and $f_3$, for an `inversion percentage' of 78\%, thus confirming the typical behavior observed in the Ellsberg three-color example.

A quantum mechanical model for our experimental data can be constructed as follows. We look for two orthogonal states $p_{w_1}$ and $p_{w_2}$, represented by the unit vectors $|w_1\rangle$ and $|w_2\rangle$, respectively, such that
\begin{eqnarray}
\langle w_1|{\mathscr F}_{1}-{\mathscr F}_{2}|w_1\rangle=0.68 \label{data1} \\
\langle w_2|{\mathscr F}_{4}-{\mathscr F}_{3}|w_2\rangle=0.69 \label{data2}
\end{eqnarray} 
where ${\mathscr F}_{i}$, $i=1,2,3,4$, have been defined in (\ref{f1})--(\ref{f4}). In the color basis $\{|R\rangle, |Y\rangle,|B\rangle\}$, which coincides with the canonical basis of ${\mathbb C}^{3}$, $|w_1\rangle$ and $|w_2\rangle$ can be written as
\begin{eqnarray}
|w_1 \rangle=(\frac{1}{\sqrt{3}} e^{i \theta_{R}},\rho_Y e^{i \theta_{Y}}, \rho_B e^{i \theta_{B}})  \\
|w_2 \rangle=(\frac{1}{\sqrt{3}} e^{i \phi_{R}},\tau_Y e^{i \phi_{Y}}, \tau_B e^{i \phi_{B}})
\end{eqnarray}
The normalization conditions $\langle w_1|w_1\rangle=\langle w_2|w_2\rangle=1$ entail
\begin{eqnarray}
\rho_Y^2+\rho_B^2=\frac{2}{3}=\tau_Y^2+\tau_B^2 \label{norm}
\end{eqnarray}
while the orthogonality condition $\langle w_1|w_2\rangle=0$ entails
\begin{eqnarray}
\frac{1}{3}\cos(\phi_{R}-\theta_{R})+\rho_{Y}\tau_{Y}\cos(\phi_{Y}-\theta_{Y})+\rho_{B}\tau_{B}\cos(\phi_{B}-\theta_{B})=0 \label{orth1} \\
\frac{1}{3}\sin(\phi_{R}-\theta_{R})+\rho_{Y}\tau_{Y}\sin(\phi_{Y}-\theta_{Y})+\rho_{B}\tau_{B}\sin(\phi_{B}-\theta_{B})=0 \label{orth2}
\end{eqnarray}
From (\ref{f1}), (\ref{f2}) and (\ref{data1}) we get
\begin{equation}
\rho_{B}=\sqrt{\frac{1}{3}-\frac{0.68}{u(100)-u(0)}}
\end{equation}
whence, by using (\ref{norm})
\begin{equation}
\rho_{Y}=\sqrt{\frac{1}{3}+\frac{0.68}{u(100)-u(0)}}
\end{equation}
Analogously, from (\ref{f3}), (\ref{f4}) and (\ref{data2}) we get
\begin{equation}
\tau_{B}=\sqrt{\frac{1}{3}+\frac{0.69}{u(100)-u(0)}}
\end{equation}
whence, by using (\ref{norm})
\begin{equation}
\tau_{Y}=\sqrt{\frac{1}{3}-\frac{0.69}{u(100)-u(0)}}
\end{equation}
The symmetry of the equations now suggests to set $\theta_{R}=\phi_{R}=0$ and $\phi_{Y}-\theta_{Y}=\phi_{B}-\theta_{B}$. This means that we have to solve a system of 11 equations in 11 unknown variables
$\theta_R$,  $\theta_Y$, $\theta_B$, $\phi_R$,  $\phi_Y$, $\phi_B$, $\rho_Y$, $\rho_B$,$\tau_Y$, $\tau_B$, and $u(100)-u(0)$.

One can verify that a solution exists for
\begin{eqnarray}
&&\theta_{R}=0 \qquad  \theta_{Y}=28^{\circ} \qquad \theta_{B}=9.3^{\circ} \\
&&\phi_{R}=0 \qquad  \phi_{Y}=208^{\circ} \qquad \phi_{B}=189.3^{\circ} \\
&&\rho_{Y}=0.787 \qquad \rho_{B}=0.216 \\ 
&&\tau_{Y}=0.206 \qquad \tau_{B}=0.790 \\
&&u(100)-u(0)=2.373
\end{eqnarray}
Hence, the states $p_{w_1}$ and $p_{w_2}$ reproducing the data collected in the Ellsberg three-color example are represented by the unit vectors
\begin{eqnarray}
|w_1 \rangle&=&(\frac{1}{\sqrt{3}}, 0.787 e^{i 28^{\circ}}, 0.216 e^{i 9.3^{\circ}})  \\
|w_2 \rangle&=&(\frac{1}{\sqrt{3}},0.206 e^{i 208^{\circ}}, 0.790 e^{i 189.3^{\circ}})
\end{eqnarray}
These states describe the conceptual Ellsberg entity with on average 55.8 yellow balls and 4.2 black balls, when the cognitive context induced by pondering about the bet between acts $f_1$ and $f_2$ is present, and 3.8 yellow balls and 56.2 black balls, when the cognitive context induced by pondering about the bet between acts $f_3$ and $f_4$ is present, for an utility $u(100)-u(0)\approx 2.4$.

This completes the construction of a quantum mechanical model for the data collected on the Ellsberg three-color example. We will see, however, that this case study can be naturally extended to more complex situations.

Let us now come to the `Machina reflection example' and represent it by using the quantum mechanical formalism. We start from the reflection example with lower tail shifts. The conceptual Machina entity is the urn with 10 red or yellow balls and 10 black or green balls, in both cases in unknown proportion.  A possible state $p_v$ of the Machina entity is represented by the unit vector $|v\rangle$ of the complex Hilbert space ${\mathbb C}^{4}$ over complex numbers. We denote by $(1,0,0,0)$, $(0,1,0,0)$, $(0,0,1,0)$ and $(0,0,0,1)$ the unit vectors of the canonical basis of ${\mathbb C}^{4}$.

Then, we consider a `color measurement' $e$ performed in a state $p_v$. The measurement has four outcomes $o_{R}$, $o_{Y}$, $o_{B}$ and $o_{G}$ corresponding to the four colors red, yellow, black and green, respectively, and it is represented by the self-adjoint operator ${\mathscr E}$ with spectral family $\{ P_{i}=|i\rangle\langle i | \}_{i=R,Y,B,G}$, where $|R\rangle$, $|Y\rangle$, $|B\rangle$ and $|G\rangle$ are the eigenvectors associated with the eigenvalues $o_{R}$, $o_{Y}$, $o_{B}$ and $o_{G}$, respectively, of ${\mathscr E}$. Equivalently, we have ${\mathscr E}=\sum_{i=1}^{4} o_{i}P_{i}$. We set $|R\rangle=(1,0,0,0)$, $|Y\rangle=(0,1,0,0)$, $|B\rangle=(0,0,1,0)$ and $|G\rangle=(0,0,0,1)$, as in the Ellsberg case. Hence,  a state $p_v$ is represented by the unit vector
\begin{equation}
|v \rangle=\rho_R e^{i \theta_{R}}|R\rangle+\rho_Y e^{i \theta_{Y}}|Y\rangle+\rho_B e^{i \theta_{B}}|B\rangle+\rho_G e^{i \theta_{G}}|G\rangle=(\rho_R e^{i \theta_{R}}, \rho_Y e^{i \theta_{Y}}, \rho_B e^{i \theta_{B}},  \rho_G e^{i \theta_{G}})
\end{equation}
By using quantum probabilistic rules, the probability $\mu_{v}(o_{i}, e)$ of drawing a ball of color $i$, $i=R,Y,B,G$, when the Machina entity is in a state $p_v$ is given by
\begin{equation}
\mu_{v}(o_{i}, e)=\langle v | P_{i} | v \rangle=|\langle i | v \rangle|^2=\rho_{i}^{2}
\end{equation}
The reflection with lower tail shifts situation requires that $\rho_{R}^{2}+\rho_{Y}^2=1/2=\rho_{B}^{2}+\rho_{G}^2$. Therefore, 
 a state $p_v$ of the Machina entity is represented by the unit vector 
\begin{equation} \label{machinastate}
|v \rangle=(\rho_R e^{i \theta_{R}},\sqrt{\frac{1}{2}-\rho_R^2} e^{i \theta_{Y}}, \rho_B e^{i \theta_{B}}, \sqrt{\frac{1}{2}-\rho_B^2} e^{i \theta_{G}})
\end{equation}
As in the Ellsberg case, let us suppose that the 
ground state $p_0$ of the conceptual Machina entity completely reflects the symmetry between the different colors. Thus, $p_0$ is represented by the unit vector $|v_0\rangle=\frac{1}{2}(1,1,1,1)$. Whenever the decision-maker is presented with the Machina paradox situation, his or her pondering about the choices to make gives rise to a cognitive context, which changes the state of the conceptual Machina entity from $p_0$ to a generally different state $p_{w}$, represented by the unit vector $|w\rangle$, as in (\ref{machinastate}). In this framework, the pondering about a choice between $f_1$ and $f_2$ will make the ground state $p_0$ of the conceptual Machina entity change to a state $p_{w_1}$ that is generally different from the state $p_{w_2}$ in which the Machina conceptual entity changes from the ground state $p_0$ in a 
pondering about the choice between $f_3$ and $f_4$. 

Let us now come to the representation of the acts $f_1$, $f_2$, $f_3$ and $f_4$ in Table 2, Section \ref{ellsbergmachina}. For a given utility function $u$, to be determined experimentally, we respectively associate $f_1$, $f_2$, $f_3$ and $f_4$ with the self-adjoint `utility operators'
\begin{eqnarray}
{\mathscr F}_{1}&=&u(0)P_R+u(50)P_Y+u(25)P_B+u(25)P_G \label{f1mach}\\
{\mathscr F}_{2}&=&u(0)P_R+u(25)P_Y+u(50)P_B+u(25)P_G \label{f2mach}\\
{\mathscr F}_{3}&=&u(25)P_R+u(50)P_Y+u(25)P_B+u(0)P_G \label{f3mach} \\
{\mathscr F}_{4}&=&u(25)P_R+u(25)P_Y+u(50)P_B+u((0)P_G \label{f4mach}
\end{eqnarray}
The corresponding expected utilities in a state $p_{v}$ are 
\begin{eqnarray}
W_{v}(f_1)&=&\langle v| {\mathscr F}_{1}|v\rangle=u(0)\rho_R^2+u(50)\rho_Y^2+\frac{1}{2}u(25) \label{w1mach}\\
W_{v}(f_2)&=&\langle v| {\mathscr F}_{2}|v\rangle=u(0)\rho_R^2+u(25)\rho_Y^2+u(50)\rho_B^2+u(25)\rho_G^2 \label{w2mach} \\
W_{v}(f_3)&=&\langle v| {\mathscr F}_{2}|v\rangle=u(25)\rho_R^2+u(50)\rho_Y^2+u(25)\rho_B^2+u(0)\rho_G^2 \label{w3mach}\\
W_{v}(f_4)&=&\langle v| {\mathscr F}_{4}|v\rangle=\frac{1}{2}u(25)+u(50)\rho_B^2+u(0)\rho_G^2 \label{w4mach}
\end{eqnarray}
All expected utilities depend on the state $p_v$, thus it is possible to find a state $p_{w_1}$ such that $W_{w_1}(f_1) > W_{w_1}(f_2)$ ($W_{w_1}(f_2) > W_{w_1}(f_1)$), and a state $p_{w_2}$ such that $W_{w_2}(f_4) > W_{w_2}(f_3)$ ($W_{w_2}(f_3) > W_{w_2}(f_4)$).  Indeed, let us consider the state $p_{YG}$ describing the conceptual situation where there are no red balls and no black balls. This state is represented by the unit vector 
\begin{equation} \label{noredblackMachina}
|v_{YG} \rangle=(0,\sqrt{\frac{1}{2}} e^{i \theta_{Y}}, 0, \sqrt{\frac{1}{2}} e^{i \theta_{G}})
\end{equation}
Then, let us consider the state $p_{RB}$ describing the conceptual situation where there are no yellow balls and no green balls. This state is represented by the unit vector 
 \begin{equation} \label{noyellowgreenMachina}
|v_{RB} \rangle=(\sqrt{\frac{1}{2}} e^{i \theta_{R}},0, \sqrt{\frac{1}{2}} e^{i \theta_{B}}, 0)
\end{equation}
By using (\ref{w1mach}), (\ref{w2mach}),  (\ref{w3mach}) and (\ref{w4mach}), we have $W_{YG}(f_1)=W_{RB}(f_4)$ and  $W_{YG}(f_2)=W_{RB}(f_3)$. Therefore, the states $p_{YG}$ and $p_{RB}$ perfectly reproduce informational symmetry in the Machina reflection example with lower tail shifts.

Let us now apply this quantum mechanical model for the Machina paradox situation to model the data collected in (L'Haridon \& Placido, 2010) on the reflection example with lower tail shifts. These authors asked 94 students to rank the four acts  in Table 2, Section \ref{ellsbergmachina}. The students' response was that 11 students preferred acts $f_1$ and $f_3$, 
44 students preferred acts $f_1$ and $f_4$, 
15 students preferred acts $f_2$ and $f_4$, and
24 students preferred acts $f_2$ and $f_3$. This entails that $44+24=68$ students over 94 reversed their preferences, for a ratio of 0.72, thus violating SEUT and in agreement with Machina's expectations (L'Haridon \& Placido, 2010). Equivalently,  
a rate of 0.59 preferred act $f_1$ over act $f_2$, and a rate of 0.63 preferred act $f_4$ over $f_3$.

A quantum mechanical model for the experimental data above can be constructed as follows. We look for two orthogonal states $p_{w_1}$ and $p_{w_2}$, represented by the unit vectors $|w_1\rangle$ and $|w_2\rangle$, respectively, such that
\begin{eqnarray}
\langle w_1|{\mathscr F}_{1}-{\mathscr F}_{2}|w_1\rangle=0.59 \label{data1mach} \\
\langle w_2|{\mathscr F}_{4}-{\mathscr F}_{3}|w_2\rangle=0.63 \label{data2mach}
\end{eqnarray} 
where ${\mathscr F}_{i}$, $i=1,2,3,4$, have been defined in (\ref{f1mach})--(\ref{f4mach}). In the color basis $\{|R\rangle, |Y\rangle,|B\rangle, |G\rangle\}$, which coincides with the canonical basis of ${\mathbb C}^{4}$, $|w_1\rangle$ and $|w_2\rangle$ can be written as
\begin{eqnarray}
|w_1 \rangle=(\rho_R e^{i \theta_{R}},\sqrt{\frac{1}{2}-\rho_R^2} e^{i \theta_{Y}}, \rho_B e^{i \theta_{B}}, \sqrt{\frac{1}{2}-\rho_B^2} e^{i \theta_{G}})  \\
|w_2 \rangle=(\tau_R e^{i \phi_{R}},\sqrt{\frac{1}{2}-\tau_R^2} e^{i \phi_{Y}}, \tau_B e^{i \phi_{B}}, \sqrt{\frac{1}{2}-\tau_B^2} e^{i \phi_{G}})
\end{eqnarray}
The normalization conditions $\langle w_1|w_1\rangle=\langle w_2|w_2\rangle=1$ does not provide further information on the unknown variables, while the orthogonality condition $\langle w_1|w_2\rangle=0$ entails
\begin{eqnarray}
\rho_{R}\tau_{R}\cos(\phi_{R}-\theta_{R})+\rho_{Y}\tau_{Y}\cos(\phi_{Y}-\theta_{Y})+\rho_{B}\tau_{B}\cos(\phi_{B}-\theta_{B})+\rho_{G}\tau_{G}\cos(\phi_{G}-\theta_{G})=0 \label{orth1mach} \\
\rho_{R}\tau_{R}\sin(\phi_{R}-\theta_{R})+\rho_{Y}\tau_{Y}\sin(\phi_{Y}-\theta_{Y})+\rho_{B}\tau_{B}\sin(\phi_{B}-\theta_{B})+\rho_{G}\tau_{G}\sin(\phi_{G}-\theta_{G})=0 \label{orth2mach}
\end{eqnarray}
Equation $\rho_R^2+\rho_Y^2=1/2$ entails
\begin{equation}
\rho_{R}=\sqrt{\frac{1}{2}-\rho_{Y}^{2}}
\end{equation}
Further we get, from (\ref{f1mach}), (\ref{f2mach}) and (\ref{data1mach}),
\begin{equation}
\rho_{B}=\sqrt{\rho_{Y}^2-\frac{0.59}{u(50)-u(25)}}
\end{equation}
whence
\begin{equation}
\rho_{G}=\sqrt{\frac{1}{2}-\rho_{Y}^2+\frac{0.59}{u(50)-u(25)}}
\end{equation}
by using $\rho_B^2+\rho_G^2=1/2$. Analogously, from (\ref{f3mach}), (\ref{f4mach}) and (\ref{data2mach}) we get
\begin{equation}
\tau_{B}=\sqrt{\tau_{Y}^{2}+\frac{0.63}{u(50)-u(25)}}
\end{equation}
whence
\begin{equation}
\tau_{G}=\sqrt{\frac{1}{2}-\tau_{Y}^{2}-\frac{0.63}{u(50)-u(25)}}
\end{equation}
by using $\tau_B^2+\tau_G^2=1/2$. Finally, one has
\begin{equation}
\tau_{R}=\sqrt{\frac{1}{2}-\tau_{Y}^{2}}
\end{equation}
One can verify that a solution exists for
\begin{eqnarray}
\theta_{R}=87.1^{\circ} \qquad  \theta_{Y}=1.6^{\circ} \qquad \theta_{B}=1.0^{\circ}  \qquad \theta_{G}=185.2^{\circ} \\
\phi_{R}=0.7^{\circ} \qquad  \phi_{Y}=191.8^{\circ} \qquad \phi_{B}=2.9^{\circ}  \qquad \phi_{G}=7.4^{\circ} \\
\rho_{R}=0 \qquad \rho_{Y}=0.71 \qquad \rho_{B}=0.38 \qquad \rho_{G}=0.60  \\ 
\tau_{R}=0.71 \qquad \tau_{Y}=0.05 \qquad \tau_{B}=0.62 \qquad \tau_{G}=0.34  \\ 
u(50)-u(25)=1.636
\end{eqnarray}
Hence, the states $p_{w_1}$ and $p_{w_2}$ reproducing the data collected in the Machina reflection example with lower tail shifts are represented by the unit vectors
\begin{eqnarray}
|w_1 \rangle&=&(0, 0.71 e^{i 1.6^{\circ}}, 0.38 e^{i 1^{\circ}}, 0.60 e^{i 185.2^{\circ}})  \\
|w_2 \rangle&=&(0.71 e^{i 0.7^{\circ}},0.05 e^{i 191.8^{\circ}}, 0.62 e^{i 2.9^{\circ}}, 0.34 e^{i 7.4^{\circ}})
\end{eqnarray}
This completes the construction of a quantum mechanical model for the Machina reflection example with lower tail shifts.

L'Haridon and Placido (2010) also tested the reflection example with upper tail shifts. The authors asked the same 94 students to rank the four acts  in Table 3, Section \ref{ellsbergmachina}. The students' response was that
8 students preferred acts $f_1$ and $f_3$, 
47 students preferred acts $f_1$ and $f_4$, 
6 students preferred acts $f_2$ and $f_4$, and
33 students preferred acts $f_2$ and $f_3$. This entails that $47+33=80$ students over 94 reversed their preferences, for a ratio of 0.85, thus violating SEUT and in agreement with Machina's expectations. Equivalently,  
a rate of 0.59 preferred act $f_1$ over act $f_2$, and a rate of 0.56 preferred act $f_4$ over $f_3$.

A quantum mechanical model for these experimental data can be constructed by following the general lines above. One can verify that the following set of parameters constitutes a solution.
\begin{eqnarray}
\theta_{R}=0.3^{\circ} \qquad  \theta_{Y}=11.6^{\circ} \qquad \theta_{B}=1.3^{\circ}  \qquad \theta_{G}=196.5^{\circ} \\
\phi_{R}=0.7^{\circ} \qquad  \theta_{Y}=193.5^{\circ} \qquad \theta_{B}=1.7^{\circ}  \qquad \theta_{G}=16.9^{\circ} \\
\rho_{R}=0.02 \qquad \rho_{Y}=0.71 \qquad \rho_{B}=0.38 \qquad \rho_{G}=0.60  \\ 
\tau_{R}=0.71 \qquad \tau_{Y}=0 \qquad \tau_{B}=0.59 \qquad \tau_{G}=0.39  \\ 
u(50)-u(25)=1.636
\end{eqnarray}
The states $p_{w_1}$ and $p_{w_2}$ reproducing the data collected in the Machina reflection example with upper tail shifts are represented by the unit vectors
\begin{eqnarray}
|w_1 \rangle&=&(0.02 e^{i 0.3^{\circ}}, 0.71 e^{i 11.6^{\circ}}, 0.38 e^{i 1.3^{\circ}}, 0.60 e^{i 196.5^{\circ}})  \\
|w_2 \rangle&=&(0.71 e^{i 0.7^{\circ}},0, 0.59 e^{i 1.7^{\circ}}, 0.39 e^{i 16.9^{\circ}})
\end{eqnarray}
Hence, also the reflection example with upper tail shifts can be modeled within our quantum theoretical framework.

\section{Toward a quantum probabilistic expected utility theory\label{QEUT}}
The analysis of the Ellsberg and Machina paradox situations in Section \ref{quantummodels} allows us to point out two fundamental aspects of our quantum theoretical approach to model human decisions under uncertainty, as follows.

(i) Our model mathematically captures ambiguity in its generality, and it is flexible enough to reproduce realistic situations where ambiguity aversion prevails, but also situations where different subjective attitudes toward ambiguity, such as ambiguity attraction, or even other behaviors, prevail. Further, a quantum mechanical approach of this type is also compatible with experimental situations where human preferences accord with the predictions of SEUT and the Sure-Thing principle holds. This provides a theoretical support to a conjecture we made in some previous papers (Aerts, Sozzo \& Tapia, 2012; 2014), where we hypothesized that ambiguity aversion is only one -- albeit a very important one -- of the cognitive aspects driving human decisions in uncertainty situations, and that, more generally, it is the overall cognitive landscape which influences the human judgment in these situations.

(ii) The mathematical element that captures ambiguity is the state describing the conceptual entity that is the object of the decision. It is this conceptual state that incorporates the subjective preferences toward ambiguity  in a very specific way, namely the ambiguity preference -- aversion, attraction or neutrality -- introduces a cognitive context at the moment when the pondering about a choice takes place, and this cognitive context changes the state of the conceptual entity under consideration -- the Ellsberg entity or the Machina entity. If we now extend our quantum theoretical modeling to any conceptual entity, then we can represent this conceptual state by means of a unit vector of a Hilbert space ${\mathscr H}$. By using Gleason theorem, we can bijectively associate this unit vector with a quantum probability distribution, the latter serving as the subjective probability distribution modeling human decisions. Therefore, one can point out a fundamental difference between SEUT and our quantum theoretical approach. Indeed, subjective probability is modeled in a single classical (Kolmogorovian) probability space over a single $\sigma$--algebra of events, according to SEUT. On the contrary, subjective probability is modeled in a quantum probability framework over a (non-Boolean) lattice of projection operators representing events, according to our quantum approach. In addition, subjective probabilities are fixed in SEUT and are the same also when situations of pondering about choices between different acts are considered. On the contrary, subjective probabilities are `state-dependent' in our approach, hence the quantum probability distribution changes together with the state of the conceptual entity when situations of pondering about different choices are considered.

Points (i) and (ii) suggest us to sketch a state-dependent generalization of SEUT based on the mathematical formalism of quantum theory and modeling human preferences. Let us call `quantum expected utility theory' (QEUT) this generalization. For the sake of simplicity, we use the same symbols, notations and simplifications that we have introduced in Section \ref{ellsbergmachina} for SEUT.

Let us consider a conceptual entity $S$ described by a Hilbert space ${\mathscr H}$, and denote by ${\mathscr L}({\mathscr H})$ the complete orthocomplemented lattice of all projection operators over ${\mathscr H}$ (see \ref{quantum}).\footnote{It is well known that ${\mathscr L}({\mathscr H})$ is, in particular, weakly modular, but it is not distributive, hence it is non-Boolean.} In each instant the conceptual entity $S$ is in a state $p_{v}$, represented by the unit vector $|v\rangle$ belonging to ${\mathscr H}$. Because of Gleason theorem, $|v\rangle$ is bijectively associated with a quantum probability distribution $p_{v}: {\mathscr L}({\mathscr H}) \longrightarrow [0,1]$, such that, for every $P \in {\mathscr L}({\mathscr H})$, $p_{v}(P)=\langle v|P|v\rangle$. The elements of ${\mathscr L}({\mathscr H})$ represent events, in this formulation. Let us now consider a set $\{ E_1, E_2, \ldots, E_n \}$ of mutually exclusive and exhaustive events, represented by the spectral family $\{ P_i \}_{i=1,2,\ldots,n}$ of orthogonal projection operators $P_1, P_2, \ldots, P_n$, so that $\sum_{i=1}^{n}P_{i}=\mathbbmss{1}$, where $\mathbbmss{1}$ is the identity operator over ${\mathscr H}$. If the event $E_i$ is associated with the consequence $x_i$, $i=1,2,\ldots,n$, then the act $f=(E_1,x_1;E_2,x_2;\ldots;E_n,x_n)$ is represented by an `utility self-adjoint operator'
\begin{equation}
{\mathscr F}=\sum_{i=1}^{n}u(x_i)P_{i}
\end{equation}
where $u$ is a Bernoulli utility function over the set $\mathscr{X}$ of consequences. Then, the act $f=(E_1,x_1;E_2,x_2;\ldots,E_n,x_n)$ is described by an `expected utility in the state $p_{v}$' given by
\begin{equation}
W_{v}(f)=\langle v|{\mathscr F}| v \rangle=\sum_{i=1}^{n}u(x_i)\langle v| P_{i}| v\rangle= \sum_{i=1}^{n}u(x_i) p_{v}(P_{i})
\end{equation}
In our representation, the quantum probability measure $p_{v}(\cdot)$ models the subjective probability associated with a human decision on the act $f$. It is important to mention that, if the conceptual entity $S$ is in a different state $p_{w}$, represented by the unit vector $|w\rangle\in {\mathscr H}$, then the decision on the act $f$ is modeled by a different quantum probability measure $p_{w}(\cdot)$.

Let us now consider two acts the act $f=(F_1,x_1;F_2,x_2;\ldots;F_n,x_n)$ and the act $g=(G_1,y_1;G_2,y_2;\ldots;G_n,y_n)$, where the event $F_i$ is associated with the consequence $x_i$, and  the event $G_i$ is associated with the consequence $y_i$, $i=1,2,\ldots,n$. Then, we introduce a `(state-dependent) preference relation' $\succsim_{v}$ in the state $p_v$, as follows.

\bigskip
\noindent
{\bf Definition 1.}
{\it We say that `act $f$ is `preferred to act $g$ in the state $p_v$' of a conceptual entity $S$, and write $f \succsim_{v} g$ if we have $W_{v}(f) \ge W_{v}(g)$, where $W_v(f)$ and $W_{v}(g)$ are the expected utilities of acts $f$ and $g$, respectively, in the state $p_v$.}

\bigskip
\noindent
Suppose now that the mutually exclusive and exhaustive events $G_1, G_2, \ldots, G_n$ are represented by the spectral family $\{ Q_i \}_{i=1,2,\ldots,n}$ of orthogonal projection operators $Q_1, Q_2, \ldots, Q_n$, so that $\sum_{i=1}^{n}Q_{i}=\mathbbmss{1}$. Then, the weak preference relation $f \succsim_{v} g$ in Definition 1 is equivalent to the requirement
\begin{equation} \label{quantumpreference}
\sum_{i=1}^{n}\langle v| u(x_i)P_{i}-u(y_i)Q_{i}|v\rangle \ge 0
\end{equation}
where ${\mathscr F}=\sum_{i}u(x_i)P_{i}$ and ${\mathscr G}=\sum_{i}u(y_i)Q_{i}$ are the utility self-adjoint operators representing acts $f$ and $g$, respectively. Equation (\ref{quantumpreference}) generalizes the preference relations that we have found in the specific cases of the Ellsberg and Machina paradox situations in Section \ref{quantummodels}. One can then determine from experimental data the utilities $u(x_i)$, $u(y_i)$ and the unit vector $|v\rangle$ modeling concrete human preferences, as we have done in Section \ref{quantummodels}.

The previous mathematical construction allows an extension of traditional SEUT that automatically incorporates ambiguity, ambiguity aversion and other relevant attitudes toward uncertainty. It is still at a descriptive level and, indeed, a normative formulation of QEUT would require the introduction of plausible axioms bijectively connecting preference relations with state-dependent expected utilities. Moreover, it is worth investigating how traditional state-independent SEUT can be recovered from state-dependent QEUT. In particular, the mutual relationships between the weak preference relation $\succsim$ in SEUT and the weak preference relation $\succsim_{v}$ in QEUT should be deepened. We plan to devote our future research to answer these questions, because any human decision, be it economic, financial, political or medical, crucially relies on a satisfactory model of decision-making in presence of uncertainty.



\appendix

\section{Fundamentals of a quantum-theoretic modeling\label{quantum}}
We illustrate in this section how the mathematical formalism of quantum theory can be applied to model situations outside the microscopic quantum world, more specifically, in the representation of conceptual entities, like the Ellsberg or the Machina entity. We avoid in our presentation superfluous technicalities, but aim to be synthetic and rigorous at the same time.

When the quantum mechanical formalism is applied for modeling purposes, each considered entity  -- in our case a conceptual entity -- is associated with a complex Hilbert space ${\cal H}$, that is, a vector space over the field ${\mathbb C}$ of complex numbers, equipped with an inner product $\langle \cdot |  \cdot \rangle$ that maps two vectors $\langle A|$ and $|B\rangle$ onto a complex number $\langle A|B\rangle$. We denote vectors by using the bra-ket notation introduced by Paul Adrien Dirac, one of the pioneers of quantum theory (Dirac, 1958). Vectors can be `kets', denoted by $\left| A \right\rangle $, $\left| B \right\rangle$, or `bras', denoted by $\left\langle A \right|$, $\left\langle B \right|$. The inner product between the ket vectors $|A\rangle$ and $|B\rangle$, or the bra-vectors $\langle A|$ and $\langle B|$, is realized by juxtaposing the bra vector $\langle A|$ and the ket vector $|B\rangle$, and $\langle A|B\rangle$ is also called a `bra-ket', and it satisfies the following properties:

(i) $\langle A |  A \rangle \ge 0$;

(ii) $\langle A |  B \rangle=\langle B |  A \rangle^{*}$, where $\langle B |  A \rangle^{*}$ is the complex conjugate of $\langle A |  B \rangle$;

(iii) $\langle A |(z|B\rangle+t|C\rangle)=z\langle A |  B \rangle+t \langle A |  C \rangle $, for $z, t \in {\mathbb C}$,
where the sum vector $z|B\rangle+t|C\rangle$ is called a `superposition' of vectors $|B\rangle$ and $|C\rangle$ in the quantum jargon.

From (ii) and (iii) follows that inner product $\langle \cdot |  \cdot \rangle$ is linear in the ket and anti-linear in the bra, i.e. $(z\langle A|+t\langle B|)|C\rangle=z^{*}\langle A | C\rangle+t^{*}\langle B|C \rangle$.

We recall that the `absolute value' of a complex number is defined as the square root of the product of this complex number times its complex conjugate, that is, $|z|=\sqrt{z^{*}z}$. Moreover, a complex number $z$ can either be decomposed into its cartesian form $z=x+iy$, or into its polar form $z=|z|e^{i\theta}=|z|(\cos\theta+i\sin\theta)$.  As a consequence, we have $|\langle A| B\rangle|=\sqrt{\langle A|B\rangle\langle B|A\rangle}$. We define the `length' of a ket (bra) vector $|A\rangle$ ($\langle A|$) as $|| |A\rangle ||=||\langle A |||=\sqrt{\langle A |A\rangle}$. A vector of unitary length is called a `unit vector'. We say that the ket vectors $|A\rangle$ and $|B\rangle$ are `orthogonal' and write $|A\rangle \perp |B\rangle$ if $\langle A|B\rangle=0$.

We have now introduced the necessary mathematics to state the first modeling rule of quantum theory, as follows.

\medskip
\noindent{\it First quantum modeling rule:} A state $A$ of an entity -- in our case a conceptual entity -- modeled by quantum theory is represented by a ket vector $|A\rangle$ with length 1, that is $\langle A|A\rangle=1$.

\medskip
\noindent
An orthogonal projection $M$ is a linear operator on the Hilbert space, that is, a mapping $M: {\cal H} \rightarrow {\cal H}, |A\rangle \mapsto M|A\rangle$ which is self-adjoint and idempotent. The latter means that, for every $|A\rangle, |B\rangle \in {\cal H}$ and $z, t \in {\mathbb C}$, we have:

(i) $M(z|A\rangle+t|B\rangle)=zM|A\rangle+tM|B\rangle$ (linearity);

(ii) $\langle A|M|B\rangle=\langle B|M|A\rangle^{*}$ (hermiticity);

(iii) $M \cdot M=M$ (idempotency).

The identity operator $\mathbbmss{1}$ maps each vector onto itself and is a trivial orthogonal projection. We say that two orthogonal projections $M_k$ and $M_l$ are orthogonal operators if each vector contained in the range $M_k({\cal H})$ is orthogonal to each vector contained in the range $M_l({\cal H})$, and we write $M_k \perp M_l$, in this case. The orthogonality of the projection operators $M_{k}$ and $M_{l}$ can also be expressed by $M_{k}M_{l}=0$, where $0$ is the null operator. A set of orthogonal projection operators $\{M_k\ \vert k=1,\ldots,n\}$ is called a `spectral family' if all projectors are mutually orthogonal, that is, $M_k \perp M_l$ for $k \not= l$, and their sum is the identity, that is, $\sum_{k=1}^nM_k=\mathbbmss{1}$. A spectral family $\{M_k\ \vert k=1,\ldots,n\}$ identifies a self-adjoint operator ${\mathscr O}=\sum_{i=1}^{n}o_kM_k$, where $o_k$ is called `eigenvalue of ${\mathscr O}$', i.e. is a solution of the equation ${\mathscr O}|o\rangle=o_k|o\rangle$ -- the non-null vectors satisfying this equation are called `eigenvectors of ${\mathscr O}$'.

The above definitions give us the necessary mathematics to state the second modeling rule of quantum theory, as follows.

\medskip
\noindent
{\it Second quantum modeling rule:} A measurable quantity $Q$ of an entity -- in our case a conceptual entity -- modeled by quantum theory, and having a set of possible real values $\{q_1, \ldots, q_n\}$ is represented by a spectral family $\{M_k\ \vert k=1, \ldots, n\}$, equivalently, by the self-adjoint operator ${\mathscr Q}=\sum_{k=1}^{n}q_kM_k$, in the following way. If the conceptual entity is in a state represented by the vector $|A\rangle$, then the probability of obtaining the value $q_k$ in a measurement of the measurable quantity $Q$ is $\langle A|M_k|A\rangle=||M_k |A\rangle||^{2}$. This formula is called the `Born rule' in the quantum jargon. Moreover, if the value $q_k$ is actually obtained in the measurement, then the initial state is changed into a state represented by the vector
\begin{equation}
|A_k\rangle=\frac{M_k|A\rangle}{||M_k|A\rangle||}
\end{equation}
This change of state is called `collapse' in the quantum jargon.

\medskip
\noindent
We denote by ${\mathscr L}({\mathscr H})$ the set of all orthogonal projection operators over the Hilbert space  ${\mathscr H}$. ${\mathscr L}({\mathscr H})$ has the algebraic properties of a complete orthocomplemented lattice, but it is not distributive, hence ${\mathscr L}({\mathscr H})$ is not a $\sigma$-algebra. A `generalized probability measure' over  ${\mathscr L}({\mathscr H})$ is a function $p: M \in {\mathscr L}({\mathscr H}) \longmapsto p(M) \in [0,1]$, such that $p(\mathbbmss{1})=1$, and $p(\sum_{k=1}^{\infty}M_k)=\sum_{k=1}^{\infty}p(M_k)$, for any countable sequence $\{ M_k \in {\mathscr L}({\mathscr H}) \}_{k=1,2,\ldots}$ of mutually orthogonal projection operators. 

Let us now suppose that the dimension of the Hilbert space is greater than 2. Then, the Born rule establishes a connection between states and generalized probability measures, as follows. Given a state of a conceptual entity represented by the unit vector $|A\rangle \in \mathscr H$, it is possible to bijectively associate  $|A\rangle$ with a generalized probability measure $p_{A}$ over ${\mathscr L}({\mathscr H})$, such that, for every $M \in {\mathscr L}({\mathscr H})$, $p_{A}(M)=\langle A |M|A\rangle$. This is the content of the so-called `Gleason theorem'. The generalized probability measure $p_{A}$ is a `quantum probability measure' over ${\mathscr L}({\mathscr H})$. It should be noticed that the Gleason theorem admits a more general formulation. However, the presentation here is sufficient for our purposes.

\medskip
\noindent
The tensor product ${\cal H}_{A} \otimes {\cal H}_{B}$ of two Hilbert spaces ${\cal H}_{A}$ and ${\cal H}_{B}$ is the Hilbert space generated by the set $\{|A_i\rangle \otimes |B_j\rangle\}$, where $|A_i\rangle$ and $|B_j\rangle$ are vectors of ${\cal H}_{A}$ and ${\cal H}_{B}$, respectively, which means that a general vector of this tensor product is of the form $\sum_{ij}|A_i\rangle \otimes |B_j\rangle$. This gives us the necessary mathematics to introduce the third modeling rule.

\medskip
\noindent
{\it Third quantum modeling rule:} A state $C$ of a compound entity -- in our case a combination of conceptual entities -- is represented by a unit vector $|C\rangle$ of the tensor product ${\cal H}_{A} \otimes {\cal H}_{B}$ of the two Hilbert spaces ${\cal H}_{A}$ and ${\cal H}_{B}$ containing the vectors that represent the states of the component entities -- concepts.

\medskip
\noindent
The above means that we have $|C\rangle=\sum_{ij}c_{ij}|A_i\rangle \otimes |B_j\rangle$, where $|A_i\rangle$ and $|B_j\rangle$ are unit vectors of ${\cal H}_{A}$ and ${\cal H}_{B}$, respectively, and $\sum_{i,j}|c_{ij}|^{2}=1$. We say that the state $C$ represented by $|C\rangle$ is a product state if it is of the form $|A\rangle \otimes |B\rangle$ for some $|A\rangle \in {\cal H}_{A}$ and $|B\rangle \in {\cal H}_{B}$. Otherwise, $C$ is called an `entangled state'.

\medskip
\noindent
The Fock space is a specific type of Hilbert space, originally introduced in quantum field theory. For most states of a quantum field the number of identical quantum entities is not conserved but is a variable quantity. The Fock space copes with this situation in allowing its vectors to be superpositions of vectors pertaining to different sectors for fixed numbers of identical quantum entities. More explicitly, the $k$-th sector of a Fock space describes a fixed number of $k$ identical quantum entities, and it is of the form ${\cal H}\otimes \ldots \otimes{\cal H}$ of the tensor product of $k$ identical Hilbert spaces ${\cal H}$. The Fock space $F$ itself is the direct sum of all these sectors, hence
\begin{equation} \label{fockspace}
{\cal F}=\oplus_{k=1}^j\otimes_{l=1}^k{\cal H}
\end{equation}
For our modeling we have only used Fock space for the `two' and `one quantum entity' case, hence ${\cal F}={\cal H}\oplus({\cal H}\otimes{\cal H})$. This is due to considering only combinations of two concepts. The sector ${\cal H}$ is called the `sector 1', while the sector ${\cal H}\otimes{\cal H}$ is called the `sector 2'. A unit vector $|F\rangle \in {\cal F}$ is then written as $|F\rangle = ne^{i\gamma}|C\rangle+me^{i\delta}(|A\rangle\otimes|B\rangle)$, where $|A\rangle, |B\rangle$ and $|C\rangle$ are unit vectors of ${\cal H}$, and such that $n^2+m^2=1$. For combinations of $j$ concepts, the general form of Fock space expressed in Equation (\ref{fockspace}) will have to be used.

This quantum-theoretic modeling can be generalized by allowing states to be represented by the so called `density operators' and measurements to be represented by the so called `positive operator valued measures'. However, our representation above is sufficient for attaining the results in this paper.

\section*{References}
\begin{description}





\item Aerts, D. (2009a). Quantum structure in cognition. {\it Journal of Mathematical Psychology 53}, 314--348.




\item Aerts, D., Broekaert, J., Gabora, L., \& Sozzo, S. (2013). Quantum structure and human thought. {Behavioral and Brain Sciences 36}, 274--276.



\item Aerts, D., \& Gabora, L. (2005a). A theory of concepts and their combinations I: The structure of the sets of contexts and properties. {\it Kybernetes 34}, 167--191.

\item Aerts, D., \& Gabora, L. (2005b). A theory of concepts and their combinations II: A Hilbert space representation. {\it Kybernetes 34}, 192--221.

\item Aerts, D., Gabora, L., \& S. Sozzo, S. (2013). Concepts and their dynamics: A quantum--theoretic modeling of human thought. {\it Topics in Cognitive Science 5}, 737--772.


\item Aerts, D., \& Sozzo, S. (2014). Quantum entanglement in conceptual combinations. {\it International Journal of Theoretical Physics 53}, 3587--3603.

\item Aerts, D., Sozzo, S., \& Tapia, J. (2012) A quantum model for the Ellsberg and Machina paradoxes. {\it Quantum Interaction. Lecture Notes in Computer Science 7620}, 48--59.

\item Aerts, D., Sozzo, S., \& Tapia, J. (2014). Identifying quantum structures in the Ellsberg paradox. {\it International Journal of Theoretical Physics 53}, 3666--3682.

\item Aerts, D., Sozzo, S., \& Veloz, T. (2015a, in print). Quantum structure of negation and conjunction in human thought. \emph{Frontiers in Psychology, doi: 10.3389/fpsyg.2015.01447}.

\item Aerts, D., Sozzo, S., \& Veloz, T. (2015b, accepted). A new fundamental evidence of non-classical structure in the combination of natural concepts. \emph{Philosophical Transactions of the Royal Society A}.



\item Allais, M. (1953). Le comportement de l'homme rationnel devant le risque. Critique des postulats et axiomes de l'ecole Am\'ericaine. {\it Econometrica 21}, 503--546.

\item Baillon, A., l'Haridon, O., \& Placido, L. (2011). Ambiguity models and the Machina paradoxes. {\it American Economic Review 101}, 1547--1560.

\item Binmore, K., Stewart, L., \& Voorhoeve, A. (2012). How much ambiguity aversion? Finding indifferences between Ellsberg’s risky and ambiguous bets. {\it Journal of Risk and Uncertainty 45}, 215--238.

\item Busemeyer, J. R., \& Bruza, P. D. (2012). {\it Quantum Models of Cognition and Decision}. Cambridge: Cambridge University Press.


\item Busemeyer, J. R., Pothos, E. M., Franco, R., \& Trueblood, J. S. (2011). A quantum theoretical explanation for probability judgment errors. {\it Psychological Review 118}, 193--218.

\item Camerer, C., \& Weber, M. (1992). Recent developments in modeling preferences: Uncertainty and ambiguity. {\it Journal of Risk and Uncertainty 5}, 325--370.

\item Dirac, P. A. M. (1958). {\it Quantum mechanics}, 4th ed. London: Oxford University Press.

\item Einhorn, H., \& Hogarth, R. (1986). Decision making under ambiguity. {\it Journal of Business 59} (Supplement), S225--S250.

\item Ellsberg, D.(1961). Risk, ambiguity, and the Savage axioms. {\it Quarterly Journal of Economics 75}. 643-–669.

\item Haven, E., \& Khrennikov, A. Y. (2013). {\it Quantum Social Science}. Cambridge: Cambridge University Press.

\item Fox, C. R., \& Tversky, A. (1995). Ambiguity aversion and comparative ignorance. {\it The Quarterly Journal of Economics 110}, 585--603.

\item Fox, C. R., \& Tversky, A. (1998). A belief-based account of decision under uncertainty. {\it Management Science 44}, 879--895.


\item Gilboa, I., \& Schmeidler, D. (1989). Maxmin expected utility with a non-unique prior. {\it Journal of Mathematical Economics 18}, 141--153.

\item Gleason, A. M. (1957). Measures on the closed subspaces of a Hilbert space. {\it Indiana University Mathematics Journal 6}, 885--893.

\item Hampton, J. A. (1988a). Overextension of conjunctive concepts: Evidence for a unitary model for concept typicality and class inclusion. {\it Journal of Experimental Psychology: Learning, Memory, \& Cognition 14}, 12--32.

\item Hampton, J. A. (1988b). Disjunction of natural concepts. {\it Memory \& Cognition 16}, 579--591.

\item Hansen, L., \& Sargent, T. (2001). Robust control and model uncertainty. {\it American Economic Review 91}, 60--66.

\item Klibanoff, P., Marinacci, M., \& Mukerji, S. (2005). A smooth model of decision making under ambiguity. {\it Econometrica 73}, 1849--1892.

\item Knight, F. H. (1921). {\it Risk, Uncertainty and Profit}. Boston: Houghton Mifflin.

\item Kolmogorov, A. N. (1933). {\it Grundbegriffe der Wahrscheinlichkeitrechnung}, Ergebnisse Der Mathematik; translated as {\it Foundations of Probability}. New York: Chelsea Publishing Company, 1950.

\item L'Haridon, O., \& Placido, L. (2010). Betting on Machina’s reflection example: An experiment on ambiguity. {\it Theory and Decision 69}, 375--393.


\item Machina, M. J. (2009). Risk, ambiguity, and the dark–dependence axioms. {\it American Economical Review 99}. 385–-392.

\item Machina, M. J., \& Siniscalchi, M. (2014). Ambiguity and ambiguity aversion. In Machina, M. J., \& Viscusi, K. (Eds.) {\it Handbook of the Economics of Risk and Uncertainty}, pp. 729--807. New York: Elsevier.

\item Maccheroni, F., Marinacci, M., \& Rustichini, A. (2006a). Ambiguity aversion, robustness, and the variational representation of preferences. {\it Econometrica 74}, 1447--1498.

\item McCrimmon, K., \& Larsson, S. (1979). Utility theory: Axioms versus paradoxes. In Allais, M., \& Hagen, O., (Eds.) {\it Expected Utility hHypotheses and the Allais Paradox}, pp. 27--145. Dordrecht: Reidel.


\item Morier, D. M., \& Borgida, E. (1984). The conjuction fallacy: A task specific phenomena? {\it Personality and Social Psychology Bulletin 10}, 243--252.

\item Pothos, E. M., \& Busemeyer, J. R. (2009). A quantum probability explanation for violations of `rational' decision theory. {\it Proceedings of the Royal Society B 276}, 2171--2178.

\item Pothos, E. M., \& Busemeyer, J. R. (2013). Can quantum probability provide a new direction for cognitive modeling? {\it Behavioral and Brain Sciences 36}. 255--274.


\item Savage, L. (1954). {\it The Foundations of Statistics}. New York: John Wiley \& Sons. Revised and Enlarged Edition (1972), New York: Dover Publications.

\item Schmeidler, D. (1989). Subjective probability and expected utility without additivity. {\it Econometrica 57}, 571--587.

\item Slovic, P., \& Tversky, A. (1974). Who accepts Savage’s axiom? {\it Behavioral Science 19}, 368--373.

\item Tversky, A., \& Kahneman, D. (1974). Judgment under uncertainty: Heuristics and biases. {\it Science 185}, 1124--1131.

\item Tversky, A. \& Kahneman, D. (1983). Extension versus intuitive reasoning: The conjunction fallacy in probability judgment. {\it Psychological Review 90}, 293--315.

\item Tversky, A., \& Kahneman, D. (1992). Advances in prospect theory: Cumulative representation of uncertainty. {\it Journal of Risk and Uncertainty 5}, 297--323.

\item Tversky, A., \& Shafir, E. (1992). The disjunction effect in choice under uncertainty. {\it Psychological Science 3}, 305--309.

\item von Neumann, J., \& Morgenstern, O. (1944). {\it Theory of Games and Economic Behavior}. Princeton: Princeton University Press.


\item Wang, Z., Solloway, T., Shiffrin, R. M., Busemeyer, J. R. (2014). Context effects produced by question orders reveal quantum nature of human judgments. {\it Proceedings of the National Academy of Sciences 111}, 9431--9436.

\end{description}

\end{document}